

%
%

\input harvmac.tex

\input epsf.tex
\overfullrule=0mm
\newcount\figno
\figno=0
\def\fig#1#2#3{
\par\begingroup\parindent=0pt\leftskip=1cm\rightskip=1cm\parindent=0pt
\baselineskip=11pt
\global\advance\figno by 1
\midinsert
\epsfxsize=#3
\centerline{\epsfbox{#2}}
{\bf Fig. \the\figno:} #1\par
\endinsert\endgroup\par
}
\def\figlabel#1{\xdef#1{\the\figno}}
\def\encadremath#1{\vbox{\hrule\hbox{\vrule\kern8pt\vbox{\kern8pt
\hbox{$\displaystyle #1$}\kern8pt}
\kern8pt\vrule}\hrule}}
\overfullrule=0pt
%

%

\def\*{\star}
\def\[{\left[}
\def\]{\right]}
\def\({\left(}
\def\){\right)}
\def\frac#1#2{{#1 \over #2}}

\def\2pi{\hbox{$2\pi i$}}
\def\e#1{{\rm e}^{^{\textstyle #1}}}

\def\dsl{\raise.15ex\hbox{/}\kern-.57em\partial}
\def\Dsl{\,\raise.15ex\hbox{/}\mkern-.13.5mu D}


%
\Title{LAVAL-PHY-22/97}{\vbox {
\centerline{ The $su(N)$ XX model}   }}
\centerline{Z. Maassarani
and P. Mathieu \foot{Work supported by NSERC (Canada) and FCAR
(Qu\'ebec).}
 }
\smallskip
\centerline{\it D\'epartement de
Physique, Universit\'e Laval, Qu\'ebec, Canada G1K 7P4}
\vskip.4in
\bigskip
\centerline{\bf Abstract}
\bigskip\vskip 0.45cm

The natural $su(N)$  generalization of the XX model is introduced and analyzed. 
It is defined in terms of the characterizing properties of the usual
XX model: the existence of two infinite sequences of mutually commuting
conservation laws and the existence of two infinite sequences of
mastersymmetries.  The integrability of these models, which cannot be obtained
in a degenerate limit of the $su(N)$-XXZ model, is established in two ways: by
exhibiting their  $R$ matrix and from a direct construction of the commuting
conservation laws. We then diagonalize the conserved laws by the
method of the algebraic Bethe Ansatz. The resulting spectrum 
is trivial in a certain sense; this provides another indication 
that the  XX $su(N)$ model is the natural generalization of the $su(2)$ model. 
The application of these models to the construction of an
integrable ladder, that is, an $su(N)$ version of the Hubbard model, is 
mentioned. 



\Date{08/97}

\vfill\eject

\newcount\eqnum \eqnum=1
\def\eq{ \eqno(\secsym\the\meqno) \global\advance\meqno by1 }
\def\eqlabel#1{ {\xdef#1{\secsym\the\meqno}} \eq }

\newwrite\refs
\def\startreferences{ \immediate\openout\refs=references
 \immediate\write\refs{\baselineskip=14pt \parindent=16pt \parskip=2pt} }
\startreferences

\refno=0
\def\aref#1{\global\advance\refno by1
 \immediate\write\refs{\noexpand\item{\the\refno.}#1\hfil\par}}
\def\ref#1{\aref{#1}\the\refno}
\def\refname#1{\xdef#1{\the\refno}}

\let\rw\rightarrow

\let\e\epsilon

\let\s\sigma
\let\a\alpha
\let\b\beta
\let\la\lambda
\def\H{{\cal H}}\def\B{{\cal B}}
\def\A{{\cal A}}

\font\tenmib=cmmib10
\font\sevenmib=cmmib10 at 7pt
\font\fivemib=cmmib10 at 5pt
\newfam\mibfam 

\textfont\mibfam=\tenmib
\scriptfont\mibfam=\sevenmib
\scriptscriptfont\mibfam=\fivemib
\mathchardef\alphaB="080B
\mathchardef\betaB="080C
\mathchardef\gammaB="080D
\mathchardef\deltaB="080E
\mathchardef\epsilonB="080F
\mathchardef\zetaB="0810
\mathchardef\etaB="0811
\mathchardef\thetaB="0812
\mathchardef\iotaB="0813
\mathchardef\kappaB="0814
\mathchardef\lambdaB="0815
\mathchardef\muB="0816
\mathchardef\nuB="0817
\mathchardef\xiB="0818
\mathchardef\piB="0819
\mathchardef\rhoB="081A
\mathchardef\sigmaB="081B
\mathchardef\tauB="081C
\mathchardef\upsilonB="081D
\mathchardef\phiB="081E
\mathchardef\chiB="081F
\mathchardef\psiB="0820
\mathchardef\omegaB="0821
\mathchardef\varepsilonB="0822
\mathchardef\varthetaB="0823
\mathchardef\varpiB="0824
\mathchardef\varrhoB="0825
\mathchardef\varsigmaB="0826
\mathchardef\varphiB="0827


\newsec{Introduction}

Much efforts have been devoted in recent years to the study of the XXZ model and
its Lie algebraic generalizations.  These models have displayed a very
rich mathematical structure and in particular an underlying quantum group
symmetry [\ref{M. Jimbo, Comm. Math. Phys. {\bf 102} (1986) 537; V.V. Bazhanov,
Comm. Math. Phys. {\bf 113} (1987) 471.}\refname\JB]. 
In the
$su(2)$ case, the XX model is obtained from the XXZ model
$$H^{\rm XXZ} = \sum_i [\s^x_i\s^x_{i+1} + \s^y_i\s^y_{i+1}+
\Delta \s^z_i\s^z_{i+1}]\eq$$
in the limit $\Delta=0$. 
Its integrability is a somewhat trivial fact since the model is equivalent to 
a free-fermion model. One of the practical uses  of the
XX model is rooted in its role as a building block for the Hubbard model
defined as [\ref{B.S. Shastry, Phys. Rev. Lett. {\bf 56} (1886) 1529,
2453.}\refname\Sh]:
$$H^{\rm Hub} = \sum_i [\s^x_i\s^x_{i+1} + \s^y_i\s^y_{i+1}+
\tau^x_i\tau^x_{i+1} + \tau^y_i\tau^y_{i+1}+ U\s^z_i\tau^z_{i}]\eq$$
where the $\s$'s and the $\tau$'s are two independent sets of Pauli matrices. In
the present work, we consider the integrable $su(N)$ extension of the XX model. 
In contrast to the $N=2$ case, the
$N>2$ models  cannot be obtained as a limiting case of the one-parameter
$su(N)$
$R$ matrix related to ${\cal U}_q(sl(N))$ [\JB], nor from any known multiparameter
deformation of the $su(N)$ XXX model (see e.g., [\ref{J. Perk and C.L. Schultz,
Phys. Lett. {\bf 84A} (1981) 407; A. Foerster, I. Roditi and L. Rodrigues,
Mod. Phys. Lett. {\bf A 11} (1996) 987.}\refname\Perk]).  Moreover, one may
suspect that by simply dropping, from the isotropic XXX version, the
contributions of the  generators associated with the  Cartan subalgebra will not
lead to an integrable model.  Hence, the first step amounts to a proper
definition of the XX model through characterizing properties that 
are expected to
be maintained in the generalized version. This is the subject of section 2.  
We then present the natural
$su(3)$ extension of the  $su(2)$ XX model and analyze it in some details.  In
particular, we display its $R$ matrix and the explicit expression of all its
conservation laws.  These results are  then  generalized to the $su(N)$ case.
We then diagonalize the transfer matrix by the
method of the algebraic Bethe Ansatz. The resulting spectrum 
is trivial in a certain sense; this provides another indication 
that the  XX $su(N)$ model is the natural generalization of the $su(2)$ model.
The choice of characterizing properties of the $su(N)$ XX spin
chain is further supported by a construction of the
$su(N)$ extension of the Hubbard model, whose detailed study is left to a
subsequent paper. All our results pertain to infinite or periodic  chains.

\newsec{The $su(2)$ XX model}

To formulate the problem of finding the integrable generalization of the XX model
in a well-defined way, we need some general characterization of the latter.  

At first we recall that the XX model has two infinite sequences of
conservation laws.  If we denote by
$H_n(\Delta)$ the unique XXZ conservation law whose leading term describes the
interaction of
$n$ contiguous spins, then one sequence is simply $H_n^{(+)}=H_n(0)$ (the defining
hamiltonian being
$H_2^{(+)}$). There is a second independent sequence
$\{H_n^{(-)}\}$ and all the members of the  whole set $\{ H_n^{(\pm)}\}$ commute
among themselves. The explicit form of these laws is [\ref{E.V. Gusev, Theor.
Math. Phys., {\bf 53} (1983) 1018; M. Grady, Phys. Rev {\bf
D25} (1982) 1103; H. Itoyama and  H. B. Thacker, Nucl. Phys. {\bf B320} (1989)
541.}, \ref{H. Araki, Comm. Math. Phys., {\bf 132} (1990)
155.}\refname\Ara, \ref{M. P. Grabowski and P. Mathieu, Ann. Phys. {\bf 243}
(1995) 299.}\refname\GraMa]: 
$$ \eqalign{ H^{(+)}_n &=
 \H_n^{xx} +\H_n^{yy}\quad\quad n~{\rm even}, \cr
&=
 \H_n^{xy} -\H_n^{yx}\quad\quad n~{\rm odd},
\cr
 H^{(-)}_n&=
 \H_n^{xy}-\H_n^{yx}\quad\quad n~{\rm even}, \cr &=
\H_n^{xx}+\H_n^{yy}\quad\quad n~{\rm odd}, \cr }\eqlabel\suhn$$
with
$$\H_n^{ab} = \sum_j h_{n,j}^{ab}, \qquad
\quad h_{n,j}^{ab}=
\s^a_j\s^z_{j+1}\dots \s^z_{j+n-2}\s^b_{j+n-1},\eq$$
and $n\geq 2$. To the above set of conserved charges, we can
obviously add 
$$ H^{(-)}_1= -2\sum_j\s^z_j\eq$$
(the factor $-2$ is introduced for later convenience.)

The standard XXZ $R$ matrix does not explain 
the presence of two infinite sequences of
conservation laws.  However, there exists a one-parameter deformation
of this $R$ matrix [\ref{J. Abad and M. Rios, J. Phys. {A
28} (1995) 3319.}\refname\AR] that leads to a two-parameter hamiltonian:
$$H = \sum [\cos \delta \, (\s^x_i\s^x_{i+1} + \s^y_i\s^y_{i+1}) - \sin\delta\, 
(\s^x_i\s^y_{i+1} - \s^y_i\s^x_{i+1})+\cosh
\gamma\, (\s^z_i\s^z_{i+1}) ]\eq$$ 
This reduces to the usual XXZ hamiltonian when
$\delta=0$.  However with $\gamma = i\pi/2$, we obtain a linear combination of
$H^{(+)}_2$ and $ H^{(-)}_2$ and these two parts commute as already mentioned.

The property of having two infinite sequences of
conservation laws does not provide a unique characterization of the model since
it is also shared by the more general XYh model, the anisotropic version of 
the XX model in a magnetic field.

Another characteristic property of the XX model is the existence of an infinite
number of mastersymmetries 
$B_n^{(\pm)}$ whose commutation with conservation laws generate conservation
laws (this is the definition of a mastersymmetry). $B_n^{(\pm)}$ is the first
moment of the charge
$H_n^{(\pm)}$. For instance
$$B_2^{(+)}= \sum_j [~j\; (\s^x_j\s^x_{j+1} + \s^y_j\s^y_{j+1})]\eq$$
and similarly, the lowest order mastersymmetry is simply $B_1^{(-)} = -2\sum
j\,\s^z_j$.  The precise expression for the commutator of the mastersymmetries
with the conservation laws is
$$[B_k^{(\e_1)}, H_n^{(\e_2)}] = 2 i(-1)^\phi (n-1)[ H_{n+k-1}^{(\e_1\e_2)}
+(-1)^\xi H_{|n-k|+1}^{(\e_1\e_2)}]\eqlabel\bhc$$
with $\e_i=\pm$; $\phi$ and $\xi$ are functions of $n,k,\e_i$.\foot{The explicit
expression of the phases is
$$\phi = {1\over 4}\,[(-1)^k+\e_1]\,[(-1)^n+\e_2]$$
and $$\xi = \left\{\matrix {& \e_1(-1)^k \qquad\qquad &{\rm for}~ k\leq n\cr &
{1\over 4}\,[(-1)^k+\e_1]\,[(-1)^n-\e_2] +1 &{\rm for}~ k> n\cr}\right.$$}
This formula  holds for all positive values of $n$ and $k$ with
the understanding that  $H_1^{(+)}=0$; in particular, it covers the case
$k=1$ which justifies the previous choice of normalization for $H_1^{(-)}$. 
It is thus clear that any  conservation law, even $H_1^{(-)}$, can be
obtained from
$H_2^{(+)}$ by the application of a suitable mastersymmetry.
Notice that the commutator of two mastersymmetries is also a mastersymmetry:
$$\eqalign{ [B_k^{(\e_1)}, B_n^{(\e_2)}] & = 2 i(-1)^\phi \,\left\{
(n-k)B_{n+k-1}^{(\e_1\e_2)} \right.\cr
&~ \left.+(-1)^{\xi} [(n+k-2)B_{|n-k|+1}^{(\e_1\e_2)} -({\rm
min}(k,n)-1)^2H_{|n-k|+1}^{(\e_1\e_2)}] \right\}\cr} \eqlabel\bbc$$

As a side remark, we stress that (\bhc) together with $[H_n^{(\pm)},H_2^{(+)}]=0$
imply the commutation of all the charges among themselves: 
$[H_n^{(\e_1)},H_m^{(\e_2)}]=0$.  This
is a consequence of the Jacobi identity with a recursion on $n$ (with
$m>n$).

Again, it is known that for the XYh model, there is also an infinite number
of mastersymmetries [\ref{E. Barouch and  B. Fuchssteiner, Stud. Appl. Math. {\bf
73} (1985) 221.}\refname\BarFuch,
\Ara].  However, only half of the XX mastersymmetries  ($B_2^{(+)}, B_3^{(-)},
B_4^{(-)}, \cdots$) survive the anisotropic deformation\foot{The existence of
 an extra  mastersymmetry in the XX model is  pointed out in
[\GraMa] but the observation that there are two infinite sequences  appears
to be new.} (in particular, the sets
$\{H_n^{(+)}\}$ and
$\{H_n^{(-)}\}$ are not related by the action of a mastersymmetry).  Hence, it
appears that requiring two infinite sequences of mastersymmetries provides a unique
characterization of the XX model. Most likely, this boils down to the simpler
criterion of the existence of the two mastersymmetries $B_1^{(-)}$ and
$B_2^{(+)}$.

\newsec{The $su(3)$ XX model}

\subsec{Definition of the model}

The standard form of the $su(3)$ XXZ model
 defined from the ${\cal U}_q(sl(3))$ $R$ matrix, written in terms of the
 Gell-Mann matrices,\foot{These are normalized as follows:
$$[\la^a, \la^b ] = 2if^{abc}\la^c ~, \qquad \la^a\la^b+\la^b\la^a =
\frac43\delta^{ab}+2d^{abc}\la^c$$ where $f^{abc}$ is completely antisymmetric
and $d^{abc}$ completely symmetric.}  reads
$$H= \sum_i \left[\sum_{a\not=3,8} \la^a_i\la^a_{i+1} + \cosh \gamma\,
(\la^3_i\la^3_{i+1} +\la^8_i\la^8_{i+1} ) -{1\over \sqrt 3} \sinh \gamma\,
(\la^3_i\la^8_{i+1} -\la^8_i\la^3_{i+1} )\right]\eq$$ 
There is no value of the
parameter  $\gamma$ that decouples the Cartan subalgebra generators. Therefore
the suitable form of the $su(3)$ XX hamiltonian must be guessed.  As a first
selecting condition we require the model to have an extra conservation law that
couples at most two spins (i.e., a candidate $H_2^{(-)}$). The naive guess 
$$H= \sum_i \sum_{a\not=3,8} \la^a_i\la^a_{i+1} \eq$$
does not satisfy this criterion.  However its truncated version
$$H\equiv H_2^{(+)} =  \sum_i \sum_{a\not= 1,2,3,8}
\la^a_i\la^a_{i+1}\eqlabel\sutrxx$$
does have the desired property:
$$H_2^{(-)} =  \sum_i [ \la^4_i\la^5_{i+1}-  \la^5_i\la^4_{i+1}+
\la^6_i\la^7_{i+1}- \la^7_i\la^6_{i+1}]\eq$$
commutes with $ H_2^{(+)}$.\foot{In ref [\AR] a two-parameter
$su(3)$ $R$ matrix is obtained, leading to the spin-chain hamiltonian density
$$\eqalign{ &\phantom{+}~\cos
\delta\; (\la^1_i\la^1_{i+1}+\la^2_i\la^2_{i+1}+
\la^6_i\la^6_{i+1}+\la^7_i\la^7_{i+1})\cr
& - \sin \delta\;
(\la^1_i\la^2_{i+1}-\la^2_i\la^1_{i+1}+\la^6_i\la^7_{i+1}-
\la^7_i\la^6_{i+1})\cr
& + \cos 2\delta\; (\la^4_i\la^4_{i+1}+\la^5_i\la^5_{i+1})-\sin 2\delta\;
(\la^4_i\la^5_{i+1}-\la^5_i\la^4_{i+1})\cr & +  \cosh \gamma \;
(\la^3_i\la^3_{i+1}
+\la^8_i\la^8_{i+1} ) -{1\over \sqrt 3} \sinh \gamma \; (\la^3_i\la^8_{i+1}
-\la^8_i\la^3_{i+1} )\cr}$$
The first term (proportional to $\cos \delta$) is unitarily equivalent to the
above  $ H_2^{(+)}$ density and the second term is likewise related to 
$H_2^{(-)}$.  However there is no value of the parameters that would make the
remaining terms disappear.  Our results suggest the existence of a three-parameter
deformation of the $R$ matrix of [\AR] that allows for the decoupling of the
undesired terms, leaving us with a one-parameter hamiltonian.}  The model
(\sutrxx) is thus our candidate XX
$su(3)$ model.  
Note that this particular model has already been identified as a
candidate integrable model in [\ref{K.-H. M\"utter and A. Schmitt, J. Phys. 
{\bf A 28} (1995) 2265.}\refname\MS],
picked up as one of the few
$su(3)$ models satisfying the Reshetikhin condition [\ref{Kulish P P and
Sklyanin E K 1982 in {\it Integrable Quantum Field Theories} ed J Hietarinta and
C Montonen (Lecture Notes in Physics {\bf 151}, Springer Verlag) p.
61.}\refname\KS].  This model also appears as a limiting case of the 19-vertex
model found in [\ref{M. Idzumi, T. Tokihiro and M. Arai, J. de Physique
{\bf I4} (1994) 1151.}\refname\ita]. 
In [\ref{M. P. Grabowski and P. Mathieu, J. Phys. {\bf A 28}
(1995) 4777.}\refname\GraMat] the Reshetikhin 
condition has been shown to be equivalent to
the existence of a boost operator $B$ satisfying $[[B,H],H]=0$, where if $H=\sum_i
h_{i,i+1}$ is the defining hamiltonian, $B$ is given by $B=\sum_i
i\, h_{i,i+1}$.  From [\MS] we thus already know that there exists a
mastersymmetry
$B_2^{(+)}$, which by itself is a strong signal 
of integrability: it automatically implies the existence of 
the conserved charges $
H_3^{(+)}$ and $ H_4^{(+)}$. Note further that (up to an irrelevant
multiplicative factor)
$H_2^{(-)}$ can be written as
$$ H_2^{(-)} = \sum_i\sum_{a,b\in\A} f^{8ab}\la_i^a\la_{i+1}^b\eq$$
where 
$\A=\{4,5,6,7\}$.  This directly shows that $H_2^{(-)}$ can be obtained from
$H_2^{(+)}$ by the application of the mastersymmetry $B_1^{(-)}$ associated to the
conserved charge $H_1^{(-)} = \sum_j \la^8_j$:
$$[B_1^{(-)}, H_2^{(+)}] = 2 i H_2^{(-)}\eq$$
(In contrast, the other degree-1 conservation law $\sum \la^3_i$ does not lead to a mastersymmetry: $[\,[\sum_j  j\la^3_j,H_2^{(+)}], H_2^{(+)}]\not=0$.) 

The existence of the mastersymmetries $B_1^{(-)}$ and $B_2^{(+)}$ singles out
the model (\sutrxx). This is the conjectured minimal characterization of an XX
model.  Its proper characterization, the existence of two infinite
families of mastersymmetries, is established in the following section.

\subsec{The $su(3)$ XX conservation laws}

To write down the $su(3)$ conservation laws in closed form, we introduce the
following notation: Latin indices take values in the set $\A=\{4,5,6,7\}$ and
Greek indices are elements of the complementary set $\B=\{1,2,3,8\}$; repeated
Latin (Greek) indices are understood to be summed in $\A$ ($\B$). We have found
the general expression for $H_n^{(+)}$ to be 
$$H_n^{(+)} = \left(\prod_{i=1}^{n-2}
f^{a_i\a_ia_{i+1}}\right)\sum_j \la_j^{a_1}\la_{j+1}^{\a_1}\la_{j+2}^{\a_2}\cdots
\la_{j+n-2}^{\a_{n-2}}\la_{j+n-1}^{a_{n-1}} \eqlabel\hnp$$ 
while that for $H_n^{(-)}$ reads 
$$H_n^{(-)} = f^{8a_0a_1}\left(\prod_{i=1}^{n-2}
f^{a_i\a_ia_{i+1}}\right)\sum_j \la_j^{a_0}\la_{j+1}^{\a_1}\la_{j+2}^{\a_2}\cdots
\la_{j+n-2}^{\a_{n-2}}\la_{j+n-1}^{a_{n-1}} \eqlabel\hnm$$
These expressions are the direct generalization of the $su(2)$ XX conservation
laws (i.e., with the replacements  $f^{abc}\rw\e^{abc}$, $\A\rw
\{1,2\}$, $\B\rw\{3\}$, they reduce to the
$su(2)$ charges $H_n^{(\e)}$ given by (\suhn) up to the sign factor
$(-1)^{[(n-2)/2]+n(\e+1)/2}$, where $[m]$ indicates the integer
part of $m$).  By a direct calculation, we have checked that 
$$[H_n^{(+)}, H_2^{(+)}] = [H_n^{(-)}, H_2^{(+)}] = 0\eq$$
and these conservation laws can be recursively generated from $H_2^{(+)}$:
$$\eqalign{ &[B_2^{(+)}, H_n^{(\pm)}] = (n-1)(H_{n+1}^{(\pm)}\mp2
H_{n-1}^{(\pm)})\cr
&  [B_1^{(-)}, H_n^{(+)}]= H_n^{(-)}\cr}\eq$$ 
These verifications use a number of
identities:
$$\eqalign { f^{a\a b}f^{a\a c} &= \frac32 \delta^{bc}\cr
f^{8ca}f^{a\a b} &= f^{8ba}f^{a\a c} \cr
f^{a p \a}f^{\a q b} d^{b\b a}&= f^{a \b b}f^{b q \gamma} d^{\gamma p a} \cr
f^{\a p b}f^{b q \b}f^{\b r c} d^{c s \a}&=- f^{\a q b}f^{b p \b}f^{\b s c} 
d^{c r \a}\cr}\eq
 $$
The successive application of the second identity leads to
$$\eqalign{ &f^{a_1\a_1 a_2}f^{a_2 \a_2 a_3}\cdots  f^{a_{n-2}\a_{n-2}a_{n-1}}
f^{8a_{n-1}a_n}\cr & = (-1)^{n-1}f^{a_{n-1}\a_{n-2} a_2}f^{a_2 \a_{n-3} a_3}
\cdots f^{a_{n-2}\a_{1}a_{n-1}} f^{8a_{n-1}a_1}\cr}\eq$$
required in most manipulations involving the $H_n^{(-)}$'s.
In this last identity there are $n-1$ factors $f$ and it should be  noticed
that the free indices $a_1,\a_1, \cdots \a_{n-2},a_n$ on the left hand
side  are read in the reverse order on the right hand side.

These identities also ensure the existence of the mastersymmetries $B_n^{(\pm)}$
whose commutation relations with the conservation laws have exactly the structure
(\bhc).  In particular, we have
$$[B_n^{(\pm)}, H_2^{(+)}] = H_{n+1}^{(\pm)}+2 H_{n-1}^{(\pm)}\eq$$
where as in the $su(2)$ case, if $H_{n}^{(\pm)} = \sum_j [h_{n}^{(\pm)}]_{j\,
j+1\, \cdots j+n-1}$, then $B_n^{(\pm)}=  \sum_j \, j\, [h_{n}^{(\pm)}]_{j\,
j+1\, \cdots j+n-1}$.  These relations  suffice to demonstrate the
presence of two infinite family of mastersymmetries. 

The comparison of the $su(3)$ conservation laws with those of  $su(2)$ 
suggests the existence of an analogue of the Jordan-Wigner transformation. 
However such a transformation does not have the expected locality property.  No
generalized Jordan-Wigner transformation has been actually found. 

\subsec{The $su(3)$ XX $R$ matrix}

If instead of Gell-Mann matrices, we use the matrices $E^{\alpha\beta}$ with
zeros everywhere except for a 1 at the intersection of line $\alpha$ and column
$\beta$, the hamiltonian reads
$$H_2^{(\pm)} = \sum_i\sum_{\a=1,2} [E^{3\a}_iE^{\a 3}_{i+1}\pm
E^{\a 3}_iE^{3 \a}_{i+1}]\eqlabel\efom$$ 
Note that this hamiltonian density is
unitarily equivalent to $$  \sum_i \sum_{\a=1,2}
[E^{\a\;\a+1}_iE^{\a+1\;\a}_{i+1}\pm
E^{\a+1\;\a}_iE^{\a\;\a+1}_{i+1}]\eqlabel\efomp$$ 
where only the matrices corresponding to the generator 
associated to the simple roots are seen to appear.
The unitary transformation matrix relating the two
hamiltonians is built out of the matrix 
$$U=\pmatrix{1&0&0\cr 0&0&1\cr0&1&0\cr}\eq$$  which permutes the  vectors 2 and 3
while leaving the vector 1 invariant. The unitary operator 
$U\otimes ... \otimes U$, one copy for each site of the chain, transforms 
one hamiltonian into the other. Such vector permutations also mean
that the whole construction still goes through  
for a hamiltonian $H_2^{(+)}$ (see (\sutrxx)) defined with
$a\not= 6,7,3,8$, or $a\not= 4,5,3,8$. 

The $R$ matrix leading to (\efom) is
$$\eqalign{ {\check R}(\lambda) = & \sin\lambda \sum_{{\a=1,2}}(e^{i\delta} E^{\a 3}\otimes E^{3\a} + e^{-i\delta}E^{3\a }\otimes E^{\a3})\cr &
+ \sum_{{\a=1,2}}(E^{\a \a}\otimes E^{33} + E^{33 }\otimes E^{\a\a})
+\cos \lambda \; [E^{33 }\otimes E^{33}+
\sum_{{\a, \b=1,2}}E^{\a \a}\otimes E^{\b\b} ]\cr}\eq$$
where ${\check R}= P R$, $P$ being the
permutation operator. (Note that
$ {\check R}(0)= {\rm Id}$.)  This is related to our $su(3)$ model in the sense that$${d{\check R}\over d\lambda}(0) = \sum_{{\a=1,2}}\left[ 
(E^{\a 3}\otimes E^{3\a} +
E^{3\a}\otimes E^{\a 3}) \cos\delta + i ( E^{\a 3}\otimes E^{3 \a} - E^{3\a
}\otimes E^{\a 3 }) \sin \delta\right]\eq$$ 
Evaluated at sites $i,i+1$, this last expression
is a linear combination of the densities of $H_2^{(\pm)}$.  
That $\check{R}$ is a genuine
$R$ matrix follows from the fact that it satisfies the Yang-Baxter equation:
$$\check R_{12}(\lambda -\mu)  \check R_{23}(\lambda)\check R_{12}(\mu) =
\check R_{23}(\mu)\check R_{12}(\lambda)\check R_{23}(\lambda -\mu)\eqlabel\ybec$$
This provides, in the framework of the quantum inverse scattering method (QISM),
an independent proof of integrability.  
And as in the $su(2)$ case, the presence of the residual parameter $\delta$   explains the origin of
the two sequences of conservation laws found for this model.

We now show how the $su(2)$ and $su(3)$ models and their features generalize
straightforwardly to $su(N)$.

\newsec{The $su(N)$ generalization}

Having unraveled the Lie algebraic structure of the $su(3)$ XX model in
the previous section, the generalization to the $su(N)$ case is clear.  
Writing  $\a<3$ instead of $\a=1,2$, the generalization simply amounts to
replacing $3$ by
$N$ in the previous formulae.  The $su(N)$ XX hamiltonians, for $N\geq 2$, are 
$$H_2^{(\pm)} = \sum_i\sum_{\a<N} [E^{\a N}_iE^{N\a}_{i+1}\pm
E^{N\a}_iE^{\a N}_{i+1}]\eqlabel\efN$$ and the $R$ matrix reads
$$\eqalign{ {\check R}(\lambda) = &\sum_{{\a<N}}(x E^{\a
N}\otimes E^{N\a}   + x^{-1} E^{N\a }\otimes E^{\a N})\sin\lambda\cr
&~ + \sum_{{\a<N}}(E^{\a
\a}\otimes E^{NN} + E^{NN }\otimes E^{\a\a})\cr &~
+ [E^{NN }\otimes E^{NN}+\sum_{{\a, \b<N}}E^{\a \a}\otimes E^{\b\b}
]\cos \lambda\cr}\eqlabel\rcn$$   
where $x=e^{i\delta}$.
It satisfies the regularity
property
$$\check R (0) = {\rm Id} \eq$$ the unitarity condition
$$\check R (\lambda) \check R (-\lambda) = {\rm Id} \;\cos^2\lambda \eq$$
and  the Yang-Baxter equation (\ybec) (and as already 
mentioned, this $R$ matrix
appears to be new).  In the QISM formulation, the conserved quantities can be
obtained from the transfer matrix 
$$\tau (\lambda) \equiv {\rm tr}_0 \; T(\lambda) \equiv 
{\rm tr}_0 \; L_{0M}...L_{01} \eqlabel\taut$$
(for a chain of $M$ sites)  
where $L_{0i}(\lambda)= R_{0i}(\lambda)$,  
the trace is taken over the auxiliary space 0
and $T(\lambda)$ is the monodromy matrix. The two-dimensional vertex model
has $N^2 + 2 N - 2$ non-vanishing Boltzmann weights. 
The Yang-Baxter equation ensures that two transfer matrices with 
different spectral parameters commute:
$$[\tau (\lambda) , \tau (\mu)] = 0 \eq$$
Therefore $\tau(\lambda)$ is a generating function for the conserved 
quantities of the one-dimensional spin-chain model.
The usual choice 
$$ H_{n+1} \propto \left({d^n \ln\tau (\la)\over d\lambda^n}\right)_{\la=0}
\;\;,\;\;\; n \geq 0\eq$$ 
gives the mutually commuting  conserved quantities studied above. 
For instance we find
$$\eqalign{& H_2 = H_2^{(+)}\cos\delta + i H_2^{(-)}\sin\delta\cr
& H_3 = H_3^{(+)}\cos (2\delta ) + i H_3^{(-)} \sin (2\delta ) \cr} \eq$$
where
$$\eqalign{ H^{(\pm)}_3 = & \sum_i \left[ \sum_{\a < N}\left( 
E_i^{\a N} E_{i+1}^{N N} E_{i+2}^{N\a} \mp E_i^{N\a} E_{i+1}^{N N} 
E_{i+2}^{\a N}\right) \right. \cr
& \left. +\sum_{\a , \b <N} \left(\pm E_i^{N\a} E_{i+1}^{\a\b} 
E_{i+2}^{\b N} - E_i^{\a N} E_{i+1}^{\b\a} E_{i+2}^{N\b}\right) \right]\cr }
\eq$$
Since $H_2$ and $H_3$ commute for all values of $\delta$ we find that
the four hamiltonians $H_2^{\pm}$, $H_3^{\pm}$  mutually commute, as expected.

In addition to these conserved charges, the transfer matrix formalism makes
quite transparent the fact that there are additional spin-1 conserved charges:
$$\eqalign{ & H_1^{\a\b}= \sum_j E_j^{\a\b}\qquad \a,\b<N\cr
& H_1^\a= \sum_j
(E_j^{\a\a}-E^{NN}_j) \qquad \a<N\cr}\eqlabel\mag$$ 
To prove the commutativity of these
charges with the transfer matrix, we observe that
$$[E_i^{\a\b}, L_{0i}(\la)] = -[E_0^{\a\b}, L_{0i}(\la)]\eq$$
The commutator $[H_1^{\a\b}, \tau(\la)]$ is then easily seen to vanish:  
$$\eqalign{ 
[H_1^{\a\b}, \tau(\la)] &= {\rm tr}_0\sum_{i=1}^M ( L_{0M}\cdots
L_{0\,i+1}[E_i^{\a\b}, L_{0i}(\la)]
L_{0\,i-1}\cdots L_{01})\cr 
&=
-{\rm tr}_0\sum_{i=1}^M ( L_{0M}\cdots
L_{0\,i+1}[E_0^{\a\b}, L_{0i}(\la)]
L_{0\,i-1}\cdots L_{01})\cr
&=-{\rm tr}_0 (E_0^{\a\b} L_{0M}\cdots
L_{01}) + {\rm tr}_0 ( L_{0M}\cdots
L_{01}E_0^{\a\b})=0\cr}\eq$$
A similar argument works for $H_1^\a$.  To the defining hamiltonian of the XX
model, we can thus add $N^2-2N+1$ independent magnetic-field terms without
spoiling the integrability of the model. The spin-1 charges can also
be seen as the   generators
of an  $su(N-1)\oplus u(1)$ symmetry.

In the generic $su(N)$ case, the  conservation laws
still take the form (\hnp) and (\hnm) with the following redefinition of
the sets $\A$ and $\B$: the elements of $\A$ are those corresponding 
to the $su(N)$  generators constructed from $E^{\a N}$ and $E^{N \a}$, e.g., 
$$\la^{2j}= E^{j N}+E^{N j}, \qquad \la^{2j+1}= -i( E^{j N} - E^{N j})\qquad
1\leq j<N\eq$$ while those of $\B$ correspond to the generators constructed from
$E^{NN}$ and $E^{\a\b}$, ($\a , \b < N$). The matrix $\la^8$ is replaced by
$\la^{N^2-1} = \sum_{j<N}E^{jj}-(N-1)E^{NN}$. Furthermore, the mastersymmetries
obtained for the  $su(2)$ and $su(3)$ models generalize directly to the $su(N)$
case.

We stress again that the special role played by the generators
corresponding to the index $N$ is only superficial. All summations encountered
above with a restriction `$\a$ less than $N$' can be replaced with the restriction
`$\a$ different from $\a_0$' with any fixed value of $\a_0$ between 1 and $N$. The models defined for different values
$\a_0$ are all related to each other by unitary transformations which permute 
the basis vectors (cf. the relation between the hamiltonians (\efom) and
(\efomp)).

With this equivalence relation in mind, it is now instructive to compare the XX
hamiltonians with their XXX and XXZ counterparts for $su(N)$. These
hamiltonians can be written as a sum of two pieces:
$$ H_2 = H_2^{\Delta} + H_2^{CSA}= \sum_i \sum_{\a \in \Delta^+}( \lambda^{\a}_i
\lambda^{-\a}_{i+1} + \lambda^{-\a}_i \lambda^{\a}_{i+1} )+ H_2^{CSA} \eq$$
where $\Delta^+$ is the set of {\it positive} roots of the $su(N)$ algebra
and  $H_2^{CSA}$ is bilinear in the generators of Cartan subalgebra.
The XXZ deformation of XXX corresponds to a modification of the 
$H_2^{CSA}$ part of the XXX hamiltonian.  The XX hamiltonian in contrast has no
CSA contribution {\it and} the  sum over the raising and lowering operators is
restricted to a subset of $N-1$ roots.

We now diagonalize the set of conserved quantities found earlier.

\newsec{Algebraic Bethe Ansatz}

Since all the hamiltonians $H_n^{(\pm)}$ 
mutually commute  it is possible 
to simultaneously diagonalize them. The $R$ matrix formulation provides
a powerful and elegant diagonalization procedure of the transfer
matrix $\tau (\lambda)$ and therefore of all the  hamiltonians. 
This method is known as the algebraic Bethe Ansatz (for a review see [\ref{H.J. De Vega, Int. J. Mod. Phys. {\bf A 4} (1989) 2371--2463, and 
references therein.}]\refname\vega).

The transfer matrix was defined in (\taut) as the trace over the auxiliary
space  of the monodromy matrix $T(\lambda)$. The latter is an $N$-dimensional 
matrix whose entries are operators acting on the Hilbert space $C^N\otimes...
\otimes C^N$, with a copy for every site. Some elements of the monodromy
matrix are used to create Ans\"atze for the eigenstates. The Yang-Baxter
equation (\ybec) implies the RTT-relation:
$$\check{R}(\lambda-\mu)\; T(\lambda)\otimes T(\mu) = T(\mu)\otimes  
T(\lambda)\;\check{R}(\lambda-\mu)\eqlabel\rtt$$
Written in components this equation provides the algebraic relations 
needed to find the action of the transfer matrix on the states.

We now use the following notation for the monodromy matrix:
$$T=\pmatrix{t_{11}&.&.&.&t_{1,N-1}&B_1\cr .&.&.&.&.&.\cr
.&.&.&.&.&.\cr  .&.&.&.&.&.\cr  t_{N-1,1}&.&.&.&t_{N-1,N-1}&B_{N-1}\cr
C_1&.&.&.&C_{N-1}&S\cr}\eqlabel\mono$$
With this notation  the transfer matrix becomes 
$\tau(\lambda)= S(\lambda) +\sum_{a=1}^{N-1} 
t_{aa}(\lambda)$.

It is easy to see that the
vector $||N\rangle \equiv |N...N\rangle$ 
is an eigenvector of the transfer matrix. 
The action  of the monodromy matrix 
on the vector $||N\rangle$ is given by
$$T(\lambda)\; ||N\rangle = \pmatrix{t(\la)&0&0&.&.&0&0\cr
0&t(\la)&0&0&.&0&0\cr 0&0&.&.&.&.&.\cr.&.&.&.&.&.&.\cr
0&0&0&.&.&t(\la)&0\cr C_1&C_2&.&.&.&C_{N-1}&s(\la)\cr} ||N\rangle \eq$$
where $t(\la)= a(\la)^M$, $s(\la)=b(\la)^M$, 
and $a(\la)=x^{-1} \sin \la$, $b(\la)=\cos \la$.  
We therefore take as Ansatz  eigenvectors the following linear 
combination of states:
$$|\lambda_1,..., \lambda_p\rangle =
\sum_{a_p,..., a_1=1}^{N-1} F^{a_p,..., a_1}
C_{a_1}(\lambda_1)...C_{a_p}(\lambda_p) \; ||N\rangle \eqlabel\ansa $$ 
where $p\leq M$. The parameters
$\lambda_i$ and the coefficients $F$ are to be determined. 

The equation (\rtt) gives the following relations:
$$\eqalign{& C_a(\lambda) C_b(\mu) = C_a(\mu) C_b(\lambda) = C_c(\mu) 
C_d(\lambda) P_{dc,ab}\cr
& S(\la)C_a(\mu) = f(\mu-\la) C_a(\mu) S(\la) + g(\mu-\la) C_a(\la) S(\mu)\cr
& t_{ab}(\la)C_c(\mu) = f(\la-\mu)C_d(\mu) t_{ae}(\la) P_{ed,bc} + 
g(\la-\mu) C_b(\la) t_{ac}(\mu)\cr 
& f(\la)= {x\cos \la \over  \sin \la}\;, \;\;\; 
g(\la)=-{x\over \sin \la}\;, \cr}\eq $$ where  all the indices 
belong to $\{1,...,N-1\}$ and $P$ is now the permutation
operator for  two copies of the $(N-1)$-dimensional space. 

One then applies the transfer matrix on the state $|\lambda_1,...,
\lambda_p\rangle$ and with the help of the above relations commutes 
the $t_{aa}$ and $S$ through the $C_i$'s. 
This creates two types of terms: the wanted terms which are proportional to
the original eigenvector, and the unwanted terms which are  required to
vanish. The $S$ and $\sum_a t_{aa}$ contributions to the wanted terms are
respectively
$$\eqalign{& s(\la) \prod_{j=1}^p f(\la_j - \la)\; |\lambda_1,..., 
\lambda_p\rangle\cr
& t(\la) \prod_{j=1}^p f(\la - \la_j)\; C(\la_1)\otimes...\otimes C(\la_p) 
\;\tau^{(N-1,p)}\; F\; ||N\rangle\cr}\eq $$
where the matrix $\tau^{(N-1,p)}$ is a transfer matrix for a $p$-sites
chain with $N-1$ states on each site.
In order for the $t$ contribution to give a vector proportional to 
$|\lambda_1,..., \lambda_p\rangle$ one requires $F$ to be an 
eigenvector of
$\tau^{(N-1,p)}$, with eigenvalue $\Lambda^{(N-1,p)}$.   

The unwanted terms arising from $S$ and $t$ have a structure similar to the
wanted terms but with one parameter $\la_j$ replaced with $\la$. Requiring 
these contributions to vanish one finds
$$ \tau^{(N-1,p)} F = (-1)^{p-1}
\left({b(\la_j)\over a(\la_j)}\right)^M  F\;\;,\;\;\; j=1,...,p
\eqlabel\aux$$

The resulting eigenvalue of the XX model is:
$$\Lambda^{(N,M)}(\la)= s(\la)\prod_{j=1}^p f(\la_j-\la) + t(\la) 
\left(\prod_{j=1}^p f(\la-\la_j) \right)\Lambda^{(N-1,p)}\eqlabel\lamb$$

One then has to diagonalize the transfer matrix $\tau^{(N-1,p)}$. 
The crucial points here
are that this transfer matrix is built out of the permutation operator
appearing in the commutation relations, and that the latter operator is in turn
an $R$ matrix. Thus one can, in principle,  embed the matrix  $\tau^{(N-1,p)}$
in an $su(N-1)$ XXX or XXZ model and  apply 
the foregoing procedure repeatedly until 
the dimension of the lower level space becomes equal to one; at this point
one has $\tau^{(1,p')}=1$ and the nested Ansatz closes.
Note that for $N=2$  no nesting is necessary.

However, a simplification to this procedure occurs. The  key point is that 
the transfer matrix  $\tau'\equiv \tau^{(N-1,p)}$ is not only constant, 
but also the {\it unit-shift}  operator on a periodic $p$-sites 
chain, with $N-1$ possible states on each site. 
Therefore  $\tau'$ can be written as a product of disjoint  permutation cycles; 
moreover one has $(\tau')^p={\rm Id}$.  The eigenvalues $\Lambda^{(N-1,p)}$ 
are then {\it at most}
$p^{\rm th}$ roots of unity and are highly degenerate. The dimensions of the 
cycles and their multiplicities
will depend on both $p$ and $N$. The coefficients
$F$ are the eigenvectors of $\tau'$.  For every eigenvalue $\Lambda^{(N-1,p)}$ 
one gets an eigenvalue $\Lambda^{(N,M)}(\la)$ from equation (\lamb), where
the parameters $\la_j$ are solutions of the closing equations (see (\aux)):
$$(-1)^{p-1} \left({b(\la_j)\over a(\la_j)}\right)^M=\Lambda^{(N-1,p)}
\;\;\;\; , \; j=1,...,p\eq $$
This form is the generalization of the $su(2)$ BA equations, for which
$\Lambda^{(N-1,p)}=1$ !

The vectors $||i\rangle\;,\; i=1,...,N-1$, are also eigenvectors of the 
XX transfer matrix. Some Bethe Ansatz states are built using
only one type of creation operators (for a given $i$), and the resulting 
Bethe Ansatz equations are $su(2)$
equations. 
 
We have thus completely diagonalized the transfer matrix of the XX $su(N)$
model. The nested Bethe Ansatz truncates and the spectrum is trivial in the 
above sense. This is one more indication that our generalized XX
model is the natural generalization of the $su(2)$ model. 

We now make some remarks.
The appearance of the permutation operator for the $su(N-1)$ model was 
to be expected since building the eigenstates over the vector $||N\rangle$
effectively removes the dimension corresponding to the basis 
vector $|N\rangle$. The $\check{R}$ matrix (\rcn) is then  clearly 
seen to yield $R^{(N-1,N-1)}=P$  when all terms involving $N$ are dropped.

Finally, for $N>2$, we show that the 
eigenvectors $|\lambda_1,..., \lambda_p\rangle$
are generically not eigenvectors of the magnetic field operators (\mag) studied
in section 4. Because the spectrum is degenerate 
this is not in contradiction with the fact that the magnetic 
operators commute with the conserved quantities.
Using the same methods as in section 4 we derive the following relations:
$$\eqalign{& \[ H_1^{\a},C_i(\la)\]=(1+\delta_{i,\a})\, 
C_i(\la)\;\;,\;\;\; i,\;\a < N\cr
& \[ H_1^{\a\b},C_i(\la)\]=\delta_{i,\b}\, 
C_{\a}(\la)\;\;,\;\;\; i,\;\a,\;\b < N\cr
& \[ \sum_{j=1}^M E^{NN}_j\;,\;C_i(\la)\]= 
-C_i(\la)\;\;,\;\;\; i,\;\a < N\cr}\eq $$
Therefore the Bethe states  are eigenvectors  only for 
one magnetic operator, with
$$\sum_{j=1}^M E^{NN}_j |\lambda_1,..., \lambda_p\rangle = 
(M-p) |\lambda_1,..., \lambda_p\rangle\; . \eq$$
One then shows that the Ansatz (\ansa) vanishes identically if
$p> M$.

\newsec{Conclusion}

We have  presented the natural $su(N)$ generalization of the XX model and
proved its complete integrability both directly and  via the 
quantum inverse scattering method. The hamiltonians and the other conserved charges were seen to fit in a unifying pattern.
The Bethe Ansatz diagonalization of the transfer matrix showed that the spectrum
is trivial in a certain sense. This is another indication that the 
XX $su(N)$ models we introduced are the natural generalization of the 
$su(2)$ model 

Note that the projection of these
$su(N)$ XX models onto $su(2)$ leads to integrable higher spin (i.e., $j=(N-1)/2$)
versions  of the $su(2)$ XX model, with non-quadratic defining hamiltonians. 

There are obvious extensions to this work. One is to consider the generalization
to other Lie algebras. Also, it would be
interesting to find the  multiparametric quantum-group symmetry underlying the
$su(N)$ XX model. We expect that there exists a multiparametric hamiltonian 
interpolating continuously between  the XX and XXZ  model, as happens for
$su(2)$.

An interesting application of the present analysis is the construction of 
integrable  generalizations of the Hubbard model. This model is known 
to be integrable in one dimension [\Sh]; its integrable structure is different
from  the usual spin chains models and the known  $su(2)$ model stands alone
outside an integrable hierarchy.
The $su(N)$ hamiltonian generalizing the Hubbard model is to be built from
two independent copies of the $su(N)$ XX model that couple through their
$H_1^{(-)}$ densities at the same site. As mentioned in the introduction, a
further support for the correctness of our characterizing properties of the  XX
model is rooted in the integrability of this generalized Hubbard model.  A strong
integrability indicator is the existence of a spin-3 conserved charge that
couples the two independent XX models along the $H_2^{(-)}$ density of one
model to the
$H_1^{(-)}$ density of the other, exactly as in the $su(2)$ case [\GraMa].  The
details of the analysis of this
$su(N)$ Hubbard model will be presented elsewhere.

\vskip2cm
\centerline{\bf Acknowledgment}
We thank Yvan Saint-Aubin for a useful discussion.


\bigskip \hrule \bigskip \centerline{{\bf References}}

\immediate\closeout\refs \vskip 0.5cm
  \message{References}\input references

\end